\documentclass[12pt,amsfonts]{article}
\begin{document}
\def\p {{\partial}}
\def\n {{\nu}}
\def\m {{\mu}}
\def\a {{\alpha}}
\def\bt {{\beta}}
\def\f {{\phi}}
\def\th {{\theta}}
\def\g {{\gamma}}
\def\eps {{\epsilon}}
\def\e {{\psi}}
\def\k {{\chi}}
\def\la {{\lambda}}
\def\na {{\nabla}}
\def\bn {\begin{eqnarray}}
\def\en {\end{eqnarray}}
\title{On the time evolution in totally constrained systems
with weakly vanishing Hamiltonian\footnote{e-mail:
$sami_{-}muslih$@hotmail.com}} \maketitle
\begin{center}
\author{S. I. MUSLIH\\Dept. of Physics\\ Al-Azhar University\\
Gaza, Palestine}
\end{center}
\hskip 5 cm

\begin{abstract}
The Dirac method treatment for finite dimensional singular
systems with weakly vanishing Hamiltonin leads to obtain the
equation of motion in terms of Parameter $\tau$. To obtain the
correct equations of motion one should use gauge fixing of the
form $\tau- f(t)=0$. It is shown that the canonical method leads
to describe the evolution in both standard and constrained finite
dimensional systems with weakly vanishing Hamiltonian in terms of
the physical time $t$, without using any gauge fixing conditions.
Besides the operator quantization of these systems is
investigated using the canonical method and it is shown that the
evolution of the state ${\Psi}$ with the time $t$ is described by
the Schr\"odinger equation $i \frac{\p \Psi}{\p t}=
{\hat{H}}\Psi$. The extension of this treatment to infinite
dimensional systems is given.
\end{abstract}
\newpage
\section{Introduction}

The canonical method [1-4] gives the set of Hamilton - Jacobi
partial differential equations [HJPDE] as

\bn &&H^{'}_{\a}(t_{\bt}, q_a, \frac{\p S}{\p q_a},\frac{\p S}{\p
t_a}) =0,\nonumber\\&&\a, \bt=0,n-r+1,...,n, a=1,...,n-r,\en where
\begin{equation}
H^{'}_{\a}=H_{\a}(t_{\bt}, q_a, p_a) + p_{\a},
\end{equation}
and $H_{0}$ is defined as \bn
 &&H_{0}= p_{a}w_{a}+ p_{\m} \dot{q_{a}}|_{p_{\n}=-H_{\n}}-
L(t, q_i, \dot{q_{\n}},
\dot{q_{a}}=w_a),\nonumber\\&&\m,~\n=n-r+1,...,n. \en

The equations of motion are obtained as total differential
equations in many variables as follows:

\bn
 dq_a=&&\frac{\p H^{'}_{\a}}{\p p_a}dt_{\a};\\
 dp_a=&& -\frac{\p H^{'}_{\a}}{\p q_a}dt_{\a};\\
dp_{\bt}=&& -\frac{\p H^{'}_{\a}}{\p t_{\bt}}dt_{\a};\\
 dZ=&&(-H_{\a}+ p_a \frac{\p
H^{'}_{\a}}{\p p_a})dt_{\a};\\
&&\a, \bt=0,n-r+1,...,n, a=1,...,n-r\nonumber \en where
$Z=S(t_{\a};q_a)$. The set of equations (4-7) is integrable [3,4]
if

\bn
dH^{'}_{0}=&&0,\\
dH^{'}_{\m}=&&0,  \m=n-r+1,...,n. \en If conditions (8)and (9) are
not satisfied identically, one considers them as new constraints
and again testes the consistency conditions. Hence, the canonical
formulation leads to obtain the set of canonical phase space
coordinates $q_a$ and $p_a$ as functions of $t_{\a}$, besides the
canonical action integral is obtained in terms of the canonical
coordinates.The Hamiltonians $H^{'}_{\a}$ are considered as the
infinitesimal generators of canonical transformations given by
parameters $t_{\a}$ respectively.

For the quantization of constrained systems we can use the
Dirac's method of quantization [5,6]. In this case we have
\begin{equation}
H^{'}_{\a}\Psi=0,\;\;\;\a=0,n-r+1,...,n,
\end{equation}
where $\Psi$ is the wave function. The consistency conditions are
\begin{equation}
[H'_{\m}, H'_{\n}]\Psi=0,\;\;\;\m,\n=1,...,r,
\end{equation}
where$[,]$ is the commutator. The constraints $H'_{\a}$ are
called first- class constraints if they satisfy
\begin{equation}
[H'_{\m}, H'_{\n}]=C_{\m\n}^{\g}H'_{\g}.
\end{equation}

In the case when the Hamiltonians $H'_{\m}$ satisfy
\begin{equation}
[H'_{\m}, H'_{\n}]=C_{\m\n},
\end{equation}
with $C_{\m\n}$ do not depend on $q_{i}$ and $p_{i}$, then from
(11) there arise naturally Dirac' brackets and the canonical
quantization will be performed taking Dirac's brackets into
commutators.

On the other hand, The path integral quantization is an
alternative method to perform the quantization of constrained
systems. If the system is integrable then one can solve equations
(4-6) to obtain the canonical phase-space coordinates as
\begin{equation}
q_{a}\equiv q_{a}(t, t_{\m}),\;\;\;p_{a}\equiv p_{a}(t,
t_{\m}),\;\;\m=1,...,r,
\end{equation}
then we can perform the path integral quantization using Muslih
method [7-10] with the action given by (7).

The aim of this paper is to to discuss the meaning of the time
evolution in constrained systems with vanishing Hamiltonian.

Even though systems with this property are treated in literature
[5,6,11,12], there exists ambiguity in describing the significance
of the time. For example some [13] claim that the standard
Hamiltonian dynamics expresses the evolution in terms of the
clock time $t$, while the systems with vanishing Hamiltonian
describe the evolution in terms of another parameter time $\tau$.
The canonical method [1-4] removes this ambiguity since we have
more than one Hamiltonian,which leads us to obtain the equations
of motion as total differential equations in my variables.

Now we will obtain the equations of motion for four constrained
systems with vanishing Hamiltonians and demonstrate the fact that
the dynamical variables are obtained in terms of time $(x_{0}=
t)$ without using any gauge fixing conditions.
\section{examples}

The procedure described in section $\mathbf{1}$ will be
demonstrated by the following examples:

$A$) Let us consider a system with th action integral as
\begin{equation}
S(q_{i}) =\int dt {\cal L}(q_{i}, \dot{q_{i}},
t),\;\;\;\;i=1,...,n,
\end{equation}
where $\cal L$ is a regular Lagrangian with Hessian $n$.
Parameterize the time $t\rightarrow\tau(t)$, with $\dot{\tau}
=\frac{d \tau}{dt}>0$. The velocities $\dot{q_{i}}$ may be
expressed as
\begin{equation}
\dot{q_{i}}= q_{i}^{'}\dot{\tau},
\end{equation}
where $ q_{i}^{'}$ are defined as
\begin{equation}
 q_{i}^{'}= \frac{dq_{i}}{d\tau}.
\end{equation}
Denote $t= q_{0}$ and $ q_{\m}=(q_{0}, q_{i}),\;\; \m=0,
1,...,n,$ then the action integral (15) may be written as
\begin{equation}
S(q_{\m}) =\int d\tau  \dot{t}{\cal L}(q_{\m}, \frac{
q_{i}^{'}}{t}),
\end{equation}
which is parameterization invariant since $L$ is homogeneous of
first degree in the velocities $ q_{\m}^{'}$ with $L$ given as
\begin{equation}
L(q_{\m}, \dot{q_{\m}}) = \dot{t}{\cal L}(q_{\m}, \frac{
q_{i}^{'}}{t}).
\end{equation}
The Lagrangian $L$ is now singular since its Hessian is $n$.

The generalized momenta conjugated to the generalized coordinates
$q_{\m}$ are defined as
\begin{equation}
p_{\m}^{\tau}=\frac{\p L}{\p q_{\m}^{'}}.
\end{equation}
Therefore the $i$-th component is
\begin{equation}
p_{i}^{\tau}=\frac{\p L}{\p q_{i}^{'}}= \frac{\p \cal L}{\p
\dot{q_{i}}},
\end{equation}
and the zeroth component is
\begin{equation}
p_{t}=\frac{\p L}{\p \dot{t}}= {\cal L} - \frac{\p \cal L}{\p
\dot{q_{i}}}\dot{q_{i}},
\end{equation}

Since $\cal L$ is regular, one can solve (21) for $\dot{q_{i}}$ in
terms of terms of $p_{i}$ and $\dot{q_{0}}$ as follows
\begin{equation}
\dot{q_{i}}= \dot{q_{i}}(p_{i}, \dot{q_{0}})= w_{i}.
\end{equation}
Substituting (23) in (22), one has
\begin{equation}
p_{t}= {\cal L}(q_{i}, \dot{q_{i}}(p_{i}, \dot{q_{0}})) -p_{i}
w_{i}.
\end{equation}
The primary constraint is
\begin{equation}
{H'}_{t} = p_{t} + H_{t}=0,
\end{equation}
where $H_{t}$ is defined as
\begin{equation}
H_{t}= -{\cal L}(q_{i}, w_{i}) + p_{i} w_{i}.
\end{equation}

Calculation show that the canonical Hamiltonian
\begin{equation}
H_{0} = -L(q_{0}, q_{i}, \dot{q_{0}}, w_{i}) +
p_{i}^{\tau}q_{i}^{'} + p_{t}\dot{q_{0}}\mid_{p_{t}= -H_{t}},
\end{equation}
vanishes identically.

The canonical method [1-4] leads us to obtain the set of Hamilton
-Jacobi partial differential equations as follows: \bn&&
{H'}_{0}= 0,\\
&&{H'}_{t} = p_{t} + H_{t}=0. \en

The equations of motion are obtained as total differential
equations in many variables as follows: \bn&&dq^{i}= \frac{\p
{H'}_{0}}{\p p_{i}}d\tau + \frac{\p {H'}_{t}}{\p p_{i}}dq^{0}=
\frac{\p {H'}_{t}}{\p p_{i}}dq^{0},\\
&&dp^{i}=- \frac{\p {H'}_{0}}{\p q_{i}}d\tau + \frac{\p
{H'}_{t}}{\p q_{i}}dq^{0}= -
\frac{\p {H'}_{t}}{\p q_{i}}dq^{0},\\
&&dp_{t}=- \frac{\p {H'}_{0}}{\p q_{0}}d\tau + \frac{\p
{H'}_{t}}{\p q_{0}}dq^{0}= 0. \en
Since
\begin{equation}
d{H'}_{t} = dp_{t} + H_{t},
\end{equation}
vanishes identically, this system is integrable and the canonical
phase space coordinates  $q_{i}$ and $p_{i}$ are obtained in
terms of the time $(q_{0}=t)$.

Till now we have discussed the Hamiltonian systems at the
classical level. Quantization of system (15) can be achieved by
means of path integral which we have discussed in reference [7],
or the operator methods. We will deal with the later.

For Dirac quantization [5,6] of constrained dynamical systems, one
takes the constraints equation as an operator whose action on the
allowed Hilbert space vectors is constrained to $zero$, i., e.,
${{H'}_{t}}\psi =0$, we obtain
\begin{equation}
[\hat{p_{t}}+ \hat{H_{t}}(\hat{q_{i}}, \hat{p_{i}})]\psi =0,
\end{equation}
the ordinary Scr\"odinger equation. The resulting quantum
theories are identical using either of the two-Lagrangians (15) or
(16).

$B$) As a second example, let us consider a regular system with
one-dimensional Lagrangian
\begin{equation}
l= \frac{1}{2} {\dot q}^{2} - V(q),
\end{equation}
where, ${\dot q}= \frac{d q}{dt}$. Replace the time by an
arbitrary paremetrization $t= t(\tau)$ [10,13], we obtain the two
dimensional Lagrangian
\begin{equation}
L= \dot t[\frac{1}{2}(\frac{q'}{\dot t})^{2} - V(q)],
\end{equation}
with $q'= \frac{dq}{d\tau}$ and $ \dot t= \frac{dt}{d\tau}$. The
Lagrangian $L$ is singular since its Hessian is $1$.

The generalized momenta $ p_{q}^{(\tau)}$ and $ p_{t}$
 are given by
\bn&& p_{q}^{(\tau)}=\frac{\p L}{\p (\frac{\p q}{\p \tau})}=
\frac{\p l}{\p {\dot q}}= p_{q},\\
&&p_{t}= \frac{\p L}{\p (\frac{\p t}{\p \tau})}= -
\frac{1}{2}(\frac{q'}{\dot t})^{2} -V(q). \en

Since the rank of the Hessian is one we have one primary
constraint as
\begin{equation}
{H'}_{t} = p_{t} + H_{t}=0,
\end{equation}
where $H_{t}$ is defined as
\begin{equation}
H_{t}= \frac{ p_{q}^{2}}{2} + V(q).
\end{equation}

The canonical Hamiltonian which is obtained as
\begin{equation}
H= -L + p_{q}^{(\tau)} q' - H_{t} \dot t,
\end{equation}
vanishes identically. We get the set of Hamilton-Jacobi partial
differential equations as follows; \bn&&{H'}_{t}= p_{t}+\frac{
p_{q}^{2}}{2} + V(q)=0,\\
&&H'=0
\en

The equations of motion are obtained as total differential
differential equations  as follows: \bn&&dq= \frac{\p H'}{\p
p_{q}^{(\tau)}}d\tau + \frac{\p {H'}_{t}}{\p p_{q}^{(\tau)}}dt=p_{q}dt,\\
&&dp_{q}=- \frac{\p H'}{\p q}d\tau + \frac{\p {H'}_{t}}{\p q}dt=
-\frac{dV}{dq}dt,\\
&&dp_{t}=- \frac{\p H'}{\p t}d\tau + \frac{\p {H'}_{t}}{\p t}dt=
0. \en Since \begin{equation} d{H'}_{t}= dp_{t} + p_{q}dp_{q} +
(\frac{dV}{dq})dq,
\end{equation}
vanishes identically, eqs. (44-46) are integrable and one obtains
the dynamical variables $q$ and $p_{q}$ in terms of the physical
time $t$.

Using the quantization procedure described in section
$\mathbf{1}$, we obtain
\begin{equation}
[\hat{p_{t}}+ \frac{1}{2} (\hat{p_{q}})^{2} + V(q)]\psi =0,
\end{equation}
which is the Schr\"odinger equation for linear harmonic
oscillator.
\section{Relativistic particle}

  In this section we shall use the canonical method [1-4] to obtain
the equations of motion for a relativistic particle of charge $e$
in an external electromagneticthis system  and for a free
relativistic particle.

$A$)Let us consider the action for a relativistic charged
particle in an external field as
\begin{equation}
S= \int L d\tau,
\end{equation}
where the Lagrangian $ L$ is given by
\begin{equation}
L= -[mc \sqrt{g^{\m\n}\dot{q_{\m}}\dot{q_{\n}}} +
\frac{e}{c}\dot{q_{\m}}A^{\m}],\;\;\m,\n=0, 1, 2, 3,
\end{equation}
and $g^{\m\n} = diag(1, -1, -1, -1)$. The momenta conjugated to
$\dot{q_{\m}}$ are
\begin{equation}
p_{\m}= \frac{\p L}{\p (\dot{q^{\m}})}= -
(\frac{mc\dot{q_{\m}}}{\sqrt{\dot{q}^{2}}} + \frac{e}{c} A_{\m}).
\end{equation}
Hence, the zeroth component $p_0$ and the ith component $p_{i}$
are given as \bn &&p_{0}= -
(\frac{mc\dot{q_{0}}}{\sqrt{\dot{q}^{2}}} +
\frac{e}{c} A_{0}),\\
&&p_{i}= - (\frac{mc\dot{q_{i}}}{\sqrt{\dot{q}^{2}}} +
\frac{e}{c} A_{i}). \en Since the rank of the Hessian matrix is
three, one can solve (53) to obtain the velocities $\dot{q_{i}}$
in terms of $\dot{q_{0}}, p_i$ and $A_i$ as
\begin{equation}
\dot{q_{i}}= \frac{k_{i}\dot{q_{0}}}{[|\mathbf{k_i}|^{2} + m^{2}
c^{2}]^{1/2}}= w_i,
\end{equation}
where $k_i$ is defined as
\begin{equation}
k_i= (p_i + \frac{e}{c} A_i).
\end{equation}
Substituting (54) in (52) one gets
\begin{equation}
p_{0} = - [|\mathbf{k_i}|^{2} + m^{2} c^{2}]^{1/2}- \frac{e}{c}A_0
= -H_0.
\end{equation}
Now the primary constraint is given as
\begin{equation}
H'_{0}= p_0 + [|\mathbf{k_i}|^{2} + m^{2} c^{2}]^{1/2} +
\frac{e}{c}A_0 =0.
\end{equation}

The canonical Hamiltonian (3) can be written as
\begin{equation}
H = - L + p_i w_i + p_0 \dot{q_{0}}.
\end{equation}
Explicit calculations show that $H$ vanishes identically and lead
to
\begin{equation}
H' =0.
\end{equation}
Equations (57) and (59) lead us to obtain the set of Hamilton-
Jacobi partial differential equations as follows

\bn &&H'=0,\\
&&H'_{0}= p_0 + H_0=0. \en

The equations of motion are obtained as total differential
equations as follows

\bn &&dq_i= \frac{\p H'}{\p p_i}d\tau + \frac{\p H'_{0}}{\p
p_i}dq_0,\\
&&dq_i= \frac{k_{i}}{[|\mathbf{k_i}|^{2} + m^{2}
c^{2}]^{1/2}}dq_0,\\
&&dp_i= -\frac{\p H'}{\p q_i}d\tau - \frac{\p H'_{0}}{\p
q_i}dq_0,\\
&&dp_i= -\frac{e}{c}[\frac{\p A_j}{\p
q^{i}}\frac{k_{i}}{[|\mathbf{k_i}|^{2} + m^{2} c^{2}]^{1/2}} +
\frac{\p A_0}{\p q^{i}}]dq_0,\\
&&dp_0= -\frac{\p H'}{\p q_i}d\tau - \frac{\p H'_{0}}{\p
q_0}dq_0,\\
&&dp_0= -\frac{e}{c}[\frac{\p A_j}{\p
q^{0}}\frac{k_{i}}{[|\mathbf{k_i}|^{2} + m^{2} c^{2}]^{1/2}} +
\frac{\p A_0}{\p q^{0}}]dq_0. \en

In order to have a consistent theory, one should consider the
total variation of $H'$ and $H'_{0}$. Since $H'$ vanishes
identically, we shall consider only the total variation of
$H'_{0}$. In fact
\begin{equation}
dH'_{0}= \frac{k_{i}}{[|\mathbf{k_i}|^{2} + m^{2}
c^{2}]^{1/2}}(dp_i +\frac{e}{c}dA_i) + dp_0 +\frac{e}{c}dA_0.
\end{equation}
Making use of equations (62-67), one gets
\begin{equation}
dH'_0 =0.
\end{equation}

Since the variations of $H'$ and $H'_{0}$ are identically zero,
no further constraints arise. Hence, the equations of motion are
integrable and the canonical phase space coordinates are obtained
in terms of parameter $q_0$.

We can use the quantization procedure discussed in previous
sections to obtain
\begin{equation}
[{\hat p}_{0} +  \sqrt{({\hat p}_i + \frac{e}{c} {\hat A}_i)^{2} +
m^{2}c^{2}} + \frac{e}{c}{\hat A}_0 ]\psi =0.
\end{equation}

$B$) To obtain the equations of motion for a free relativistic
particle we can use the results obtained in section $A$, but
instead we consider no interaction with external field
($A_{\m}=0$). In this case the equations of motion are obtained as
\bn &&dq_i= \frac{\p H'}{\p p_i}d\tau + \frac{\p H'_{0}}{\p
p_i}dq_0,\\
&&dq_i= \frac{p_{i}}{[|\overrightarrow{p}|^{2} + m^{2}
c^{2}]^{1/2}}dq_0,\\
&&dp_i= -\frac{\p H'}{\p q_i}d\tau - \frac{\p H'_{0}}{\p
q_i}dq_0,\\
&&dp_i= 0,\\
&&dp_0= -\frac{\p H'}{\p q_i}d\tau - \frac{\p H'_{0}}{\p
q_0}dq_0,\\
&&dp_0= 0. \en

Equations (71-76) are integrable [7]. Hence, we obtain
\begin{equation}
p_{i}= a_{i},
\end{equation}
and
\begin{equation}
p_{0}= c_{i},
\end{equation}

where $a_{i}$and $c_{i}$ are arbitrary constants. So integration
of (72) will give
\begin{equation}
q_{i}= \frac{p_{i}}{[|\overrightarrow{p}|^{2} + m^{2}
c^{2}]^{1/2}}q_0 + {\a}_{i},
\end{equation}
where ${\a}_{i}$ are constants.

Again as for the relativistic charged particle we can quantize
this system to obtain
\begin{equation}
[{\hat p}_{0} +  \sqrt{({\hat{\overrightarrow{p}}} +
m^{2}c^{2}}]\psi =0.
\end{equation}
\section{Hamilton-Jacobi treatment of the field systems}
The Hamilton-Jacobi treatment for finite dimensional systems
which are discussed in previous sections will be extended to the
field theory. Let us consider the action of this theory as
\begin{equation}
S[\phi_{A}]=\int d^{D+1}t {\cal
L}(\phi_{A},\p_{a}\phi_{A}),\;\;a=0, 1,...,D,\;\;A=1,...,N.
\end{equation}
where ${\cal L}$ is a regular Lagrangian density. Introduce $D+1$
parameters $T^{a}$, with   $det\mid{\frac{\p T^{a}}{\p t^{b}}}\mid
> 0$, and treat the $t^{a}$ on the same footing as the original
field $\phi_{A}$. The action (81) is now reads as
\begin{equation}
S[\phi_{A}, t^{a}]=\int d^{D+1}T {\cal
L}_{T}(\phi_{A},\p_{a}\phi_{A}),
\end{equation}
where \bn&& {\cal L}_{T}= {\cal L}(\phi_{A}, \frac{\p
\phi_{A}}{\p T^{a}}(J^{(-1)})_{a}^{b}) \det(J_{a}^{b}),\\
&&J_{a}^{b} = \frac{\p t^{a}}{\p T^{b}}. \en The action (82) is
invariant under parameterization $T\rightarrow{T'}(T)$. Now the
Lagrangian $ {\cal L}_{T}$ is singular since it is linear in the
velocities $\frac{\p {t}^{a}}{\p T^{0}}$.  The momenta are now
\bn&& \pi^{A}= \frac{\p {\cal L}_{T}}{\p(\frac{\p \phi_{A}}{\p
T^{0}})}=\frac{\p {\cal L}}{\p(\frac{\p
\phi_{A}}{\p t^{0}})},\\
&&\pi^{a}= \frac{\p {\cal L}_{T}}{\p(\frac{\p {t}^{a}}{\p
T^{0}})}. \en

Since ${\cal L}$ is regular, one can solve (85) for $\frac{\p
\phi_{A}}{\p T^{0}}$ in terms of $\pi^{A}, \phi_{A}, t_{a}$ and
$\frac{\p {t}^{a}}{\p T^{0}}$. Substituting (85) in (86), one has
\begin{equation}
\pi_{a} = - \frac{\p {\cal L}}{\p(\p_{a}
{\phi}_{A})}{\p_{a}}{\phi_{A}} + {\delta}_{a}^{b}{\cal L}.
\end{equation}
Hence, the primary constraint is
\begin{equation}
{H'}_{a}= \pi_{a} + H_{a}(\pi_{A}, t^{a}, \phi^{A})= 0,
\end{equation}
where
\begin{equation}
H_{a} = ( \frac{\p {\cal L}}{\p(\p_{a}
{\phi}_{A})}{\p_{a}}{\phi_{A}} - {\delta}_{a}^{b}{\cal
L})Adj(J_{b0}).
\end{equation}
Besides the canonical Hamiltonian $H_{T}$ is defined as
\begin{equation}
H_{T}= \int d^{D}T(\pi^{A}\frac{\p \phi_{A}}{\p T^{0}} +
\pi^{a}\frac{\p t^{a}}{\p T^{0}}|_{\pi_{a} = - H_{a}} -{\cal
L}_{T}).
\end{equation}
Calculations show that $H_{T}$ vanishes identically.

Now we would like to obtain the equations of motion for this
system using the canonical method [1-4]. Making use of equations
(88) and (90), one obtains the set of Hamilton-Jacobi partial
differential equations as \bn&&{H'}_{T}= 0,\\
&&{H'}_{a}= \pi_{a} + H_{a}=0. \en

The equations of motion are obtained as total differential
equations as follows: \bn&&d\phi^{A}= \frac{\p {H'}_{T}}{\p
\pi_{A}}dT^{a} + \frac{\p {H'}_{a}}{\p \pi_{A}}dt^{a}=
\frac{\p {H'}_{a}}{\p \pi_{A}}dt^{a},\\
&&d\pi^{A}=- \frac{\p {H'}_{T}}{\p \phi_{A}}dT^{a} - \frac{\p
{H'}_{a}}{\p \phi_{A}}dt^{a}= -
\frac{\p {H'}_{a}}{\p \phi_{A}}dt^{a},\\
&&d\pi_{a}=- \frac{\p {H'}_{T}}{\p t^{a}}dT^{a}- \frac{\p
{H'}_{a}}{\p t^{a}}dt^{a}= \frac{\p {H'}_{a}}{\p t^{a}}dt^{a}. \en

Since
\begin{equation}
d{H'}_{a}= d\pi_{a} + dH_{a},
\end{equation}
vanishes identically, the set of equations (93-95) is integrable
and the canonical phase space coordinates $\phi^{A}$ and $\pi^{A}$
are obtained in terms of $t^{a}$.

To obtain the path integral quantization for this system, we can
use the canonical path integral formulation discussed in
references [7-10]. Making use of (7), the canonical action
integral is calculated as
\begin{equation}
S[\phi^{A}] =\int [- H_{a} + \pi^{A}\frac{\p {H'}_{a}}{\p
\pi_{A}}]dt^{a}.
\end{equation}
Then the path integral integral for the field system is given as
\begin{equation}
\langle Out\mid S\mid In\rangle= \int
\prod_{A=1}^{N}d\phi^{A}~D\pi^{A}\exp i\int_{t^{a}}^{{t'}^{a}} [-
H_{a} + \pi^{A}\frac{\p {H'}_{a}}{\p \pi_{A}}]dt^{a}.
\end{equation}
Integrating over $\pi^{A}$ gives
\begin{equation}
\langle Out\mid S\mid In\rangle= \int
\prod_{A=1}^{N}d\phi^{A}\exp i\{\int_{t^{a}}^{{t'}^{a}}
d^{D+1}t~{\cal L}(\phi_{A}, \p_{a}\phi_{A})\}.
\end{equation}

\section{ Conclusion}

In this work we use the canonical method to investigate the
significance of time for constrained systems with weakly
vanishing Hamiltonian. For these systems we obtain the dynamical
variables in terms of physical time $t$ without fixing any gauge.

In general, the determination of the Hamiltonians ${H'}_{\a}$ is
the crucial step. If the system is integrable (Jacobi system),
then the equations of motion are obtained in terms of  $t_{\a}$.

for the first and the second example, the equations of motion are
obtained in terms of two parameters $\tau$ and $t$. However,
since the Hamiltonian ${H'}$ vanishes identically, the parameter
$\tau$ doe not appear in the equations. Hence, any parameter can
be replaced by $\tau$ and one can use $t$ as evolution parameter
without losing the Hamiltonian structure of the equations of
motion.

Besides, as for the first two examples, without fixing any gauge,
the equations of motion for a relativistic particle discussed in
section $\mathbf{3}$ are obtained in terms of parameter $q^{0}$.

In section $\mathbf{4}$ we have obtained the equations of motion
for field systems using the Hamilton-Jacobi method. In general,
continuous systems (fields) are treated as systems with infinite
degrees of freedom. The Lagrangian density is defined and the
usual variational principle technique is used to obtain the
equations of motion. the quantities $t^{a}= x, y, z, t$ are
completely independent indices of the theory. However, using the
parameterization $t^{a}\rightarrow T^{a}$ to these theories leads
to obtain the equations of motion in terms of $t^{a}$ and
$T^{a}$. Moreover, since the Hamiltonian ${H'}_{T}$ vanishes
identically, the canonical phase space coordinates $ \phi^{A}$
and $\pi^{A}$ are obtained in terms of $t^{a}$ and the parameters
$T^{a}$ do not appear in the equations of motion. On the other
hand the path integral for this system is obtained as an
integration over the canonical phase space coordinates $\phi^{A}$
and $\pi^{A}$ without using any gauge fixing conditions.

As a conclusion it is obvious that the canonical method leads to
obtain the dynamical variables for finite dimensional systems
with weakly vanishing Hamiltonian in terms of the time $t$
without using any gauge fixing conditions. The extension of this
treatment to field systems leads to obtain the equations of
motion in terms of parameters $t^{a}$ and the number of
independent parameters depend on the Hessian and the integrability
conditions.

The operator quantization for the above system is obtained using
the Hamilton Jacobi method without using any gauge fixing
conditions. In this case the evolution of the state ${\Psi}$ is
described by the Schr\"odinger equations ($i \frac{\p \Psi}{\p
t}= {\hat{H}}\Psi$) (34, 48, 70, 80). Where the operators
$\hat{H}$ are the quantum Hamiltonians corresponding to the
classical physical Hamiltonians given in equations (26, 40, 56,
80).


\begin{thebibliography}{widest-label}
\bibitem{1} Y. G\"uler, Nuovo Cimento B, 107 (1992)1389.
\bibitem{2} Y. G\"uler, Nuovo Cimento B, 107 (1992)1143.
\bibitem{3} S. I. Muslih and Y. G\"uler, Nuovo Cimento B, 110
(1995) 307.
\bibitem{4} S. I. Muslih and Y. G\"uler, Nuovo Cimento B, 113
(1998)277.
\bibitem{5}P. A. M. Dirac, {\it Lectures on Quantum Mechanics},Belfer
 Graduate School of Science, Yehiva University (A cademic Press,
 New York)1964.
\bibitem{6} P. A. M. Dirac, Can. J. Math., 2 (1950)129.
\bibitem{7}S. I. Muslih, Nuovo Cimento B, 115 (2000)1.
\bibitem{8}S. I. Muslih, Nuovo Cimento B, 115 (2000)7.
\bibitem{9}S. I. Muslih," Path integral quantization of Yang-Mills theory" submitted fo
Nuovo Cimento B.
\bibitem{10} S. I. Muslih, Hadronic J, 23 (2000)203.
\bibitem{11}A. Hanson, T. Regge and C. Teitelboim, {\it Constrained
Hamiltonian Systems} (Accademia Nazionole dei Lincei, Rome) 1976.
\bibitem{12}D. M. Gitman and I. Tyutin, {\it Quantization of Fields
with Constraints} (Springer- Verlag, Berlin, Heidelberg) 1990.
\bibitem{13}J. Goldeberg, E. Newman and C. Rovelli, J. Math. Phys.
32 (1991) 2739.


\end{thebibliography}
\end{document}